\documentclass[a4paper,twocolumn,pra,aps,nofootinbib,showpacs]{revtex4}
\newcommand{\beq}{\begin{equation}}
\newcommand{\eeq}{\end{equation}}
\newcommand{\beqa}{\begin{eqnarray}}
\newcommand{\eeqa}{\end{eqnarray}}

\def\ra{\rangle}

\usepackage{amsmath,amsfonts,amssymb}
\usepackage{graphicx}
\DeclareGraphicsExtensions{.pdf,.png,.jpg}
\usepackage{epstopdf}
\usepackage{bm}
\usepackage{dsfont}

\usepackage{color}

\DeclareMathOperator\erf{erf}

\begin{document}
\title{Fast transport of mixed-species ion chains within a Paul trap}
\author{M. Palmero$^{1}$}
\author{R. Bowler$^{2}$}
\author{J. P. Gaebler$^{2}$}
\author{D. Leibfried$^{2}$}
\author{J. G. Muga$^{1, 3}$}

\affiliation{$^{1}$Departamento de Qu\'{\i}mica F\'{\i}sica, Universidad del Pa\'{\i}s Vasco - Euskal Herriko Unibertsitatea, 
Apdo. 644, Bilbao, Spain}
\affiliation{$^{2}$National Institute of Standards and Technology, 325 Broadway, Boulder, Colorado 80305}
\affiliation{$^{3}$Department of Physics, Shanghai University, 200444 Shanghai, People's Republic of China}
\begin{abstract}
We investigate the dynamics of mixed-species ion crystals during transport between spatially distinct locations in a linear Paul trap in the diabatic regime. In a general mixed-species crystal, all degrees of freedom along the direction of transport are excited by an accelerating well, so unlike the case of same-species ions, where only the center-of-mass-mode is excited, several degrees of freedom have to be simultaneously controlled by the transport protocol. We design protocols that lead to low final excitations in the diabatic regime using invariant-based inverse-engineering for two different-species ions and also show how to extend this approach to longer mixed-species ion strings. Fast transport of mixed-species ion strings can significantly reduce the operation time in certain architectures for scalable quantum information processing with trapped ions.
\end{abstract}
\maketitle
\section{Introduction}
One possible route to scale quantum information processing  based on trapped ions \cite{Cirac}  incorporates the transport of small strings of ions between storing and processing sites \cite{Wineland,Kielpinski}.  In a recent experimental demonstration of this approach \cite{Home}, transport and subsequent sympathetic recooling of ion chains to near the ground state of motion have been among the most time consuming building blocks. Excitations might be avoided by adiabatically moving the ions, at the price 
of large transport duration and  higher susceptibility to ion heating from ambient noise fields \cite{Blakestad,noise}. In principle, it is permissible to excite the motion of the ions during transport, as long as all excitations are removed at the end of the transport \cite{review}. As we will show below, this general approach may lead to transport durations that are much shorter than what would be possible in an adiabatic approach. Previous work concentrated on transport of one particle, cold neutral atom clouds,  two ions, or ion clouds  \cite{review,Schulz,David,Huber,Calarco,Torrontegui, Li, Masuda, Uli,noise,Bowler, Walther, Palmero,Jofre,BEC}. Here, we 
study the transport of mixed-species ion chains with initial and final excitations of the motion close to the ground state. The use of two different ion species allows for sympathetic cooling of the ion motion of one species without disturbing the quantum information held by the other species \cite{Home}. Another building block utilized in \cite{Home,Hanneke} required transport of a four-ion crystal, where two ions carry the qubit information and the other two are used to cool the coupled motion of the crystal. We first study the transport of two different mass ions, and design protocols to transport them over a distance of 370 $\mu$m in durations signifficantly smaller than 100 $\mu$s
leaving them in a low energy state of motion. Our approach employs invariant based inverse engineering of shortcuts to adiabaticity \cite{Torrontegui, Palmero}. We then extend these techniques  to longer ion chains, and specifically a four-ion chain. 
We limit the study of 2- and 4-ion chains since they are enough to 
perform one- and two-qubit gates and therefore to build a universal set of gates while avoiding the problems inherent to
longer chains. 
\section{Invariant-based inverse engineering}
The invariant-based inverse-engineering method has  proved useful for single-particle transport \cite{Torrontegui,Li,Uli}, 
and  for several equal mass ions \cite{Palmero}.  For one particle of mass $m$ in 1D the Hamiltonians 
that belong to the ``Lewis-Leach family'' \cite{LL} may be written in terms of a potential $U$ that moves along $\alpha(t)$, 
and a force $F$ as  
\beq\label{Lewis}
H=\frac{p^2}{2m}-F(t)q+\frac{1}{2}m\omega ^2(t)q^2+\frac{1}{\rho^2(t)}U\left[\frac{q-\alpha(t)}{\rho(t)}\right],
\eeq
where $p$ is the momentum, $\rho$ is a scaling length parameter, and $\omega$ an angular frequency.  
This $H$ has the following dynamical invariant 
\beqa
I&=&\frac{1}{2}m[\rho (p-m\dot{\alpha})-m\dot{\rho}(q-\alpha)]^2
\nonumber\\
&+&\frac{1}{2}m\omega_0^2\left(\frac{q-\alpha}{\rho}\right)^2+U\left(\frac{q-\alpha}{\rho}\right), 
\eeqa
provided 
the functions $\rho$, $\alpha$, $F$ and $\omega$ satisfy the auxiliary equations 
\beqa
\label{auxiliary1}
\ddot{\rho}+\omega^2(t)\rho=\frac{\omega_0^2}{\rho^3},\\
\label{auxiliary2}
\ddot{\alpha}+\omega^2(t)\alpha =\frac{F(t)}{m}. 
\eeqa
For the simple case in which the potential is purely harmonic with constant angular frequency $\omega(t)=\omega_0$ we have 
$U=0$, $F(t)=m\omega_0^2Q_0(t)$, where $Q_0(t)$ is the trap trajectory;  $\alpha(t)$ becomes a classical trajectory satisfying a Newton's equation for the moving trap, and the scaling length parameter is $\rho=1$, therefore the auxiliary equation (\ref{auxiliary1}) is trivially satisfied. The inverse engineering strategy imposes boundary conditions for $\alpha$ at the boundary times $t_b=\{0,t_f\}$, where the transport starts at $t=0$ and ends at $t=t_f$. With $\alpha(0)=\dot{\alpha}(t_b)=0$, and $\alpha(t_f)=d$, the static asymptotic Hamiltonians ($H(t\le 0)$ and $H(t\ge t_f)$) and the invariant commute at the initial and final times. In this manner, the eigenstates of the initial trap are transported (mapped) via the dynamical modes of the invariant up to the eigenstates of the final trap.  In addition, $\ddot\alpha(t_b)=0$ is usually imposed to provide a continuous trap trajectory at the boundary times.  Then $\alpha(t)$ is interpolated and, by substituting $\alpha(t)$ into Eq. (\ref{auxiliary2}), we may solve for the trap trajectory $Q_0(t)$.  In general the evolution is diabatic, with transient excitations but no final excitation by construction. 
%
%
%
\section{Dynamical normal-mode coordinates\label{2ionsdm}}
Our goal is to transport a chain of ions with different mass between two sites separated by a distance $d$  in a  time $t_f$ without final motional excitation. We assume tight radial confinement so that the transport dynamics of each ion is effectively one-dimensional, and also that the external trap potential is harmonic. We label the ions as $i=1,2,...,N$. They have position coordinates $q_1,q_2,...,q_N$ and masses $m_1,m_2,...m_N$.  With the position of the minimum of the external potential $Q_0=Q_0(t)$, the Hamiltonian is  
\beq
\label{Hamiltonian}
H=\sum_{i=1}^N\frac{p_i^2}{2m_i}+\sum_{i=1}^N\frac{1}{2}u_0(q_i-Q_0)^2
+\sum_{i=1}^{N-1}\sum_{j=i+1}^N\frac{C_c}{q_i-q_j},
\eeq
where $u_0$ is the spring constant of the external trap, and $C_c=\frac{e^2}{4\pi\epsilon_0}$, with $\epsilon_0$ the vacuum permittivity.  
For later use let us also define the potential $V\equiv H-\sum_{i=1}^N\frac{p_i^2}{2m_i}$.
We assume that all ions have the same charge $e$, and that their locations obey   $q_1>q_2>\dots>q_N$, with negligible overlap of probability densities due to the strong Coulomb repulsion. For equal masses \cite{Palmero},  the dynamics for the center of mass and relative motion are uncoupled. The motion of the trap only affects the center of mass, whose dynamics is governed by a  Lewis-Leach Hamiltonian (\ref{Lewis}), so that transport without final excitation may be designed as described for a single particle.  
However, for ions with different masses, center of mass and relative motions are coupled. To cope with this coupling we apply a 
dynamical normal mode approach that approximately separates the Hamiltonian into a sum of independent harmonic oscillators. The equilibrium positions  $\{q_i^{(0)}\}$, are found by solving the system $\{\partial V/\partial q_i=0\}$ for all ions.  For $N=2$  the equilibrium positions are 
\beq
q_1^{(0)}=Q_0+x_0/2,\; q_2^{(0)}=Q_0-x_0/2,
\eeq
where 
\beq
x_0=2 \left({\frac{C_c}{4u_0}}\right)^{1/3}.
\eeq
Diagonalizing $V_{ij}=\frac{1}{\sqrt{m_im_j}}\frac{\partial^2V}{\partial q_i\partial q_j}\big|_{\{q_i,q_j\}=\{q_i^{(0)},q_j^{(0)}\}}$, we get the eigenvalues
\beq
\lambda_\pm =\omega_1^2\left[1+\frac{1}{\mu}\pm\sqrt{1-\frac{1}{\mu}+\frac{1}{\mu ^2}}\right],
\eeq
where $\omega_1=(u_0/m_1)^{1/2}$, and $\mu = m_2/m_1$, with $\mu\geq 1$. 
These eigenvalues are related to the normal-mode angular frequencies  by 
\beq
\Omega_\pm=\sqrt{\lambda_\pm}.
\eeq
The eigenvectors are 
$v_\pm=\left(\begin{array}{c}a_\pm\\b_\pm\end{array}\right)$, where 
\beqa
\label{coefficients}
a_+&=&\left(\frac{1}{1+\left(1-\frac{1}{\mu}-\sqrt{1-\frac{1}{\mu}+\frac{1}{\mu ^2}}\right)^2\mu}\right)^{1/2},
\nonumber\\
b_+&=&\left(1-\frac{1}{\mu}-\sqrt{1-\frac{1}{\mu}+\frac{1}{\mu ^2}}\right)\sqrt{\mu}a_+,
\nonumber\\
a_-&=&\left(\frac{1}{1+\left(1-\frac{1}{\mu}+\sqrt{1-\frac{1}{\mu}+\frac{1}{\mu ^2}}\right)^2\mu}\right)^{1/2},
\nonumber\\
b_-&=&\left(1-\frac{1}{\mu}+\sqrt{1-\frac{1}{\mu}+\frac{1}{\mu ^2}}\right)\sqrt{\mu}a_-.
\eeqa
Thus, the mass-weighted, dynamical, normal-mode coordinates are 
\beqa
\label{transform}
\sf{q}_+\!\!&=&\!\!a_+\sqrt{m_1}\left(q_1\!-\!Q_0\!-\!\frac{x_0}{2}\right)\!+\!b_+\sqrt{\mu m_1}\left(q_2\!-\!Q_0\!+\!\frac{x_0}{2}\right),
\nonumber\\
\sf{q}_-\!\!&=&\!\!a_-\sqrt{m_1}\left(q_1\!-\!Q_0\!-\!\frac{x_0}{2}\right)\!+\!b_-\sqrt{\mu m_1}\left(q_2\!-\!Q_0\!+\!\frac{x_0}{2}\right),
\nonumber\\
\eeqa
and the inverse transformations are
\beqa
\label{inversetransform}
q_1&=&\frac{1}{\sqrt{m_1}}\left(b_-{\sf{q}}_+-b_+\sf{q}_-\right)+Q_0+\frac{x_0}{2},
\nonumber\\
q_2&=&\frac{1}{\sqrt{\mu m_1}}\left(-a_-{\sf{q}}_++a_+\sf{q}_-\right)+Q_0-\frac{x_0}{2}.
\eeqa
Unlike the usual treatments for static traps \cite{Morigi}, we have to consider explicitly the time dependence of the 
parameter $Q_0(t)$ when writing down the Hamiltonian in the new coordinates.  
We apply the change-of-variables unitary operator
\beq
\label{unitary}
U=\int d{{\sf{q}}}_+d{{\sf{q}}}_-dq_1dq_2 |\sf{q}_+,\sf{q}_-\rangle\langle\sf{q}_+,\sf{q}_-| q_1,q_2\rangle\langle q_1,q_2 |,
\eeq
where the transformation matrix is  
\beq
\langle {\sf{q}}_+,{\sf{q}}_-|q_1,q_2\rangle =\delta [q_1-q_1({\sf{q}}_+,{\sf{q}}_-)]\delta [q_2-q_2({\sf{q}}_+,{\sf{q}}_-)].
\nonumber
\eeq
The Hamiltonian in the 
new frame is $H'=UHU^\dagger -i\hbar U(\partial_t U^\dagger)$, and the wavefunction $|\psi'\ra=U|\psi\ra$.  
For the part $UHU^\dagger$ we substitute the definitions (\ref{inversetransform}) 
in the Hamiltonian (\ref{Hamiltonian}) for $N=2$. For the non-inertial term, $-i\hbar U(\partial_t U^\dagger)$, we apply the chain rule in Eq. (\ref{inversetransform}) and Eq. (\ref{transform}). Keeping only terms up to the harmonic approximation, 
\beqa
UHU^\dagger &=&\frac{{\sf{p}}_+^2}{2}+\frac{1}{2}\Omega _+^2{\sf{q}}_+^2+\frac{{\sf{p}}_-^2}{2}+\frac{1}{2}\Omega _-^2
\sf{q}_-^2,
\nonumber\\
-i\hbar U(\partial_t U^\dagger)&=&-P_{0+}{\sf{p}}_+-P_{0-}{\sf{p}}_-,
\eeqa
where $\sf{p}_\pm$ are momenta conjugate to $\sf{q}_\pm$, and  
\beq
P_{0\pm}=\dot{Q}_0(\sqrt{m_1}a_\pm+\sqrt{\mu m_1}b_\pm).
\eeq
The linear-in-momentum terms are cumbersome for a numerical or analytical treatment, so we apply a further transformation to the frame moving with the center of the trap and remove them formally \cite{Sara}.  The wave function is transformed as $|\psi''\ra=\mathcal{U}|\psi'\ra$, whereas the corresponding Hamiltonian takes the form  $H ''=\mathcal{U} H'\mathcal{U}^\dagger+i\hbar (\partial_t \mathcal{U})\mathcal{U}^\dagger$. 
We choose   
$\mathcal{U}=e^{-i(P_{0+}{\sf{q}}_+ +P_{0-}{\sf{q}}_-)/\hbar}$ to shift the momenta, 
so that, each mode Hamiltonian in  
\beqa
H''&=&\frac{{\sf{p}}_+^2}{2}+\frac{1}{2}\Omega_+^2 \left({\sf{q}}_++\frac{\dot{P}_{0+}}{\Omega_+^2}\right)^2
\nonumber\\
&+&\frac{{\sf{p}}_-^2}{2}+\frac{1}{2}\Omega_-^2 \left({\sf{q}}_-+\frac{\dot{P}_{0-}}{\Omega_-^2}\right)^2
\label{hdprime}
\eeqa
belongs to the Lewis-Leach family.
\section{Inverse engineering for two modes}
\begin{figure}[t!]
\begin{center}
\includegraphics[width=7.5cm]{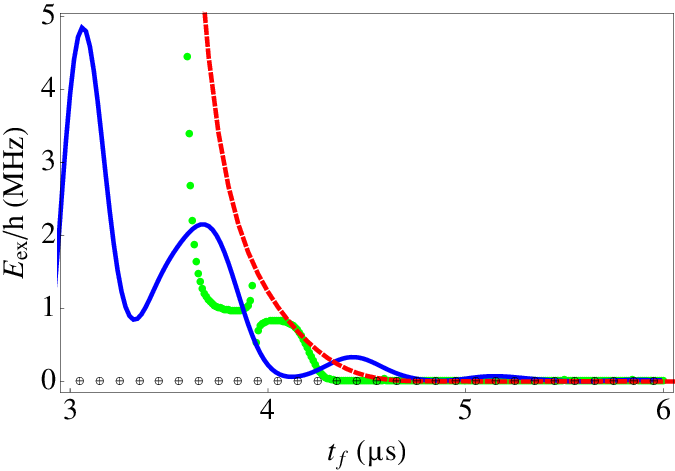}
\caption{\label{limite}(Color online) 
Motional excitation quanta vs. transport duration $t_f$ for the two ions, transported over $d=370$ $\mu$m using the exact Hamiltonian. The external potential minimum moves according to the nonic polynomial $Q_0(t;\{a_n\})$ set to satisfy Eq. (\ref{auxiliarymodes}) (green dots); the polynomial ansatz trajectory $Q_0(t;\{b_n\})$, Eq.  (\ref{goodtrajectory}), (blue-solid line); and the cosine ansatz trajectory $Q_0(t;\{c_n\})$, Eq. (\ref{costrajectory}), (red-dashed line).  The excitation for the nonic polynomial trajectory $Q_0(t;\{a_n\})$ using the uncoupled Hamiltonian (\ref{hdprime}) is also shown (black symbols). The parameters used are $\omega_1 /(2\pi)=2$ MHz, masses of $^9$Be$^+$  for the first ion and $^{24}$Mg$^+$ for the second. Both ions are initially in the motional ground state.}  
\end{center}
\end{figure}

\begin{figure}[t!]
\begin{center}
\includegraphics[width=4.2cm]{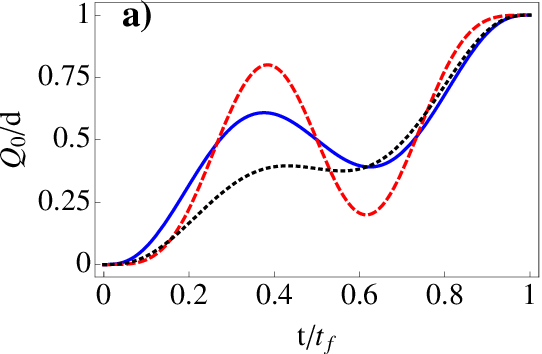}
\includegraphics[width=4.2cm]{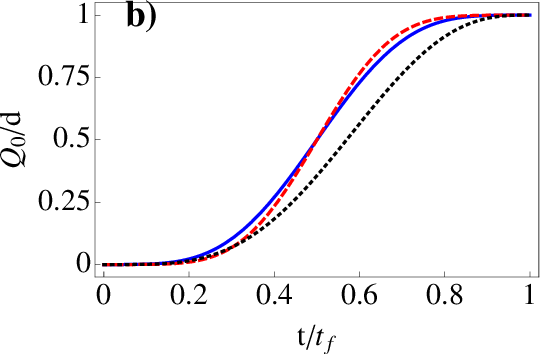}
\caption{\label{trayectorias}(Color online) 
Trap trajectories given by $Q_0(t;\{a_n\})$ (black-dashed line), Eq. (\ref{goodtrajectory}) (blue-solid line), and Eq. (\ref{costrajectory}) (red-dashed line) for different final times. 
a) $t_f=2\pi/\omega_1$;
b) $t_f=10 \times 2\pi/\omega_1$.
$\omega_1 /(2\pi)=2$ MHz, masses of $^9$Be$^+$
for the first ion and $^{24}$Mg$^+$ for the second, $d=370$ $\mu$m.}  
\end{center}
\end{figure}

The invariants corresponding to the Hamiltonians in Eq. (\ref{hdprime}) 
are known and the trajectory can be designed to avoid excitations. We also impose $\dot{Q}_0({t_b})(0)=0$ so that  $|\psi''(0)\ra=|\psi'(0)\ra$ and $|\psi''(t_f)\ra=|\psi'(t_f)\ra$. Primed and double-primed wave functions are related to each other by the unitary transformation in such a way that their initial and final states coincide. The auxiliary equations analogous to Eq. (\ref{auxiliary2}) for the modes in  Eq. (\ref{hdprime}) are  
\beq
\label{auxiliarymodes}
\ddot{\alpha}_\pm+\Omega_\pm^2\alpha_\pm=-\dot{P}_{0\pm}, 
\eeq
where the $\alpha_\pm$ are the centers of  invariant-mode wavefunctions  in the doubly-primed space \cite{Torrontegui}. Now, we can design these $\alpha_\pm$ functions to get unexcited modes after the transport, and from them inverse engineer $\dot{P}_{0\pm}$. We set the boundary conditions 
\beq
\alpha_\pm(t_b)=\dot{\alpha}_\pm(t_b)=\ddot{\alpha}_\pm(t_b)=0.
\eeq
Substituting these conditions into Eq. (\ref{auxiliarymodes}), we find $\ddot{Q}_0(t_b)=0$ for both modes. To satisfy all the conditions in Eq. (\ref{auxiliarymodes}), we try a polynomial ansatz $Q_0(t;\{a_n\})=\sum_{n=0}^9a_nt^n$. We fix $a_{0-5}$ as functions of $a_{6-9}$ so that $Q_0(0)=0$, $Q_0(t_f)=d$, $\dot{Q}_0(t_b)=\ddot{Q}_0(t_b)=0$. We then select the solutions $\alpha_\pm$ in Eq. (\ref{auxiliarymodes}) 
that satisfy $\alpha_\pm(t_b)=0$, which implies $\ddot{\alpha}_\pm(t_b)=0$, since $\dot{P}_{0,\pm}(t_b)=0$ in Eq. (\ref{auxiliarymodes}). The four parameters $a_{6-9}$ are calculated numerically for each $t_f$ by solving the system of four equations $\dot{\alpha}_\pm(t_b)=0$. Fig. \ref{limite} shows that, for the approximate  Hamiltonian with two uncoupled modes, the final excitation vanishes (see the black-symbols horizontal line). However, the higher order terms in the actual Hamiltonian modify and couple the modes, exciting the system at short transport times (green dots in Fig. \ref{limite}). 
The approach we have just described requires a numerical evaluation of the coefficients to find $Q_0(t;\{a_n(t_f)\})$ for each $t_f$. Therefore, we considered a different approximation that yields an analytical solution $Q_0(t)$ with $Q_0(0)=0$, $Q_0(t_f)=d$, $\dot{Q}_0(t_b)=\ddot{Q}_0(t_b)=0$. The resulting  $Q_0(t)$ leads to a similar level of final excitation when inserted into the full Hamiltonian as the more accurate approach. We first rewrite the Hamiltonian (\ref{Hamiltonian}) in the center of mass, $Q~=~(m_1/M)q_1+(m_2/M)q_2$, and relative, $r=q_1-q_2$, coordinates, with $M=m_1+m_2$, 
\beqa
\label{cmhamiltonian}
H&=&\frac{P^2}{2M}+\frac{1}{2}M\omega^2(Q-Q_0)^2
\nonumber\\
&+&\frac{p^2}{2m_r}+\frac{1}{2}m_r\omega _r^2r^2+\frac{C_c}{r}
\nonumber\\
&+&\frac{m_2-m_1}{2}\omega^2(Q-Q_0)r,
\eeqa
where $m_r =m_1m_2/M$, $\omega^2 =2 u_0/M$, $\omega_r^2=(m_1^2+m_2^2)/(2m_1m_2)\omega^2$, and $P$ 
is the total momentum. Neglecting the coupling term  in (\ref{cmhamiltonian}), we can construct  trap trajectories that leave the center of mass unexcited. Rewriting $\alpha=Q_c$, we first design $Q_c$ and then obtain $Q_0$ from Eq. (\ref{auxiliary2}). The four boundary conditions $\dot{Q}_0(t_b)=\ddot{Q}_0(t_b)=0$
are consistent with $Q_c^{(3)}(t_b)=Q_c^{(4)}(t_b)=0$ along with the conditions $Q_c(0)=0$, $Q_c(t_f)=d$, $\dot{Q}_c(t_b)=\ddot{Q}_c(t_b)=0$. 
We assume a polynomial ansatz $Q_c(t)=d\sum_{n=0}^9 b_ns^n$ that satisfies all conditions and obtain
$Q_0(t)$ from Eq. (\ref{auxiliary2}),  
\beq
\label{goodtrajectory}
Q_0(t)=\frac{d}{t_f^2\omega^2}\sum_{n=0}^9 b_nn(n-1)s^{n-2}+d\sum_{n=0}^9 b_ns^n,
\eeq
where $s=t/t_f$ and $\{b_0,...,b_9\}=\{0,0,0,0,0,126,-420,540,-315,70\}$ for {\it all} values of $t_f$. An alternative ansatz  with a sum of Fourier-cosines also leads to analytical expressions, 
\beqa
\label{costrajectory}
Q_c(t)&=&\frac{d}{256}\left\{c_0+\sum_{n=1}^{3}c_n\cos\left[\frac{(2n-1)\pi t}{t_f}\right]\right\},
\nonumber\\
Q_0(t)&=&\frac{d\pi^2}{256\omega ^2t_f^2}\sum_{i=1}^3 -c_n(2n-1)^2\cos\left[\frac{(2n-1)\pi t}{t_f}\right]
\nonumber\\
&+&\frac{d}{256}\left\{c_0+\sum_{n=1}^{3}c_n\cos\left[\frac{(2n-1)\pi t}{t_f}\right]\right\},
\eeqa
where $\{c_0,...,c_3\}=\{128,-150,25,-3\}$.
The resulting trap trajectories (\ref{goodtrajectory}), (\ref{costrajectory}) are simple and explicit and lead to small excitations in a similar range of parameters as the approach based on normal-modes. Some example trajectories for different transport durations are shown in Fig. \ref{trayectorias}.
\section{Four and $N$ ions}
\begin{figure}[t!]
\begin{center}
\includegraphics[width=7.5cm]{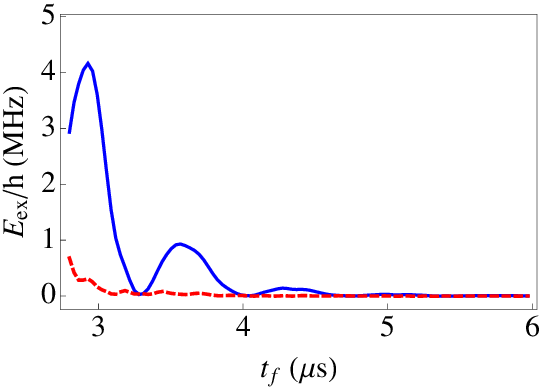}
\caption{\label{exci4}(Color online) 
Final excitation energy for a Be-Mg-Mg-Be chain transported over $d=370$ $\mu$m using the external potential minimum trajectory  in Eq. (16) with  $\omega= \sqrt{4 u_0/M}$(blue solid line) and with  $\omega=0.983 \sqrt{4 u_0/M}$ (red dashed line). The calculation is based on classical equations of motion with the ions at rest in their equilibrium positions at $t=0$.}  
\end{center}
\end{figure}

We  extend now the  normal-mode approach to $N$-ion chains, with dynamical normal mode coordinates 
\beq
{\sf{q}}_\nu=\sum_{j=1}^Na_{\nu j}\sqrt{m_j}(q_j-\delta_j^{(0)}-Q_0),
\eeq
and corresponding momenta $p_\nu$, where the
equilibrium points with respect to the trap center, $\delta_j^{(0)}$, are in general found numerically. 
Generalizing Eq. (\ref{hdprime}) to $N$ ions we find the uncoupled normal-mode Hamiltonian 
\beq
H''=\sum_{\nu=1}^N\frac{\sf{p}_\nu^2}{2}+\sum_{\nu=1}^N\frac{1}{2}\Omega_\nu^2\left({\sf{q}}_\nu+\frac{\dot{P}_{0\nu}}{\Omega_\nu^2}\right)^2,
\eeq
where $P_{0\nu}=\dot{Q}_0\sum_j a_{\nu j}m_j^{1/2}$, and $\Omega_\nu$ is the angular frequency of the $\nu$-th normal mode. 
The auxiliary equations that have to be satisfied for all $\nu$ simultaneously are
\beq
\ddot{\alpha}_\nu+\Omega_\nu^2\alpha_\nu=-\dot{P}_{0\nu}.
\eeq
Further imposing, in analogy to Eq. (\ref{auxiliarymodes}), $\alpha_\nu(t_b)=\dot{\alpha}_\nu(t_b)=\ddot{\alpha}_\nu(t_b)=0$ implies $\dot{Q}_0(t_b)=\ddot{Q}_0(t_b)=0$, exactly as for $N=2$. Thus we may construct approximate trap trajectories that are in fact identical in form to the ones for $N=2$ in Eqs.  (\ref{goodtrajectory}) or (\ref{costrajectory}), but with $\omega=\sqrt{N u_0/M}$. We find that the final excitations for a four-ion Be-Mg-Mg-Be chain (see blue solid line in Fig. \ref{exci4}),  are very similar to those for Be-Mg shown in Fig. \ref{limite}.  We can improve the results even further by treating $\omega$ as a variational free parameter. The red dashed line in Fig. \ref{exci4} shows the final excitation for $\omega=0.983 \sqrt{4 u_0/M}$.   
The calculations for the 4-ion chain are performed with classical trajectories for the ions, initially at rest in their equilibrium positions. The corresponding quantum calculation is very demanding, but it is not expected to deviate significantly from the classical result \cite{Palmero} in the nearly harmonic regime considered here.  
{For transporting longer ion chains longer final times will be needed, as more non-harmonic terms and couplings terms 
would be neglected in the normal-mode
approximation.} 
\section{Discussion}
The approximate approaches we have implemented to transport ions of different mass 
without final excitation may be compared with other approaches: the ``compensating force approach'' \cite{Masuda,Torrontegui}, the transport based on 
a linear-in-time displacement of the trap
{or a more refined error-function trajectory \cite{Reichle}.}   

Let us first discuss the ``compensating force approach'' \cite{Masuda,Torrontegui}. 
The idea behind  is that the acceleration of the trap induces in the trap frame  a non-inertial Hamiltonian term $MQ\ddot{Q_0}(t)$, $M$ being the total mass of the ion chain  and $Q$ the center of mass coordinate, that may be exactly 
compensated by applying a time-dependent term $H_{\rm{com}}=-MQ\ddot{Q_0}(t)$.  This has been discussed for $N$-equal  masses \cite{Masuda2,Palmero,Adol} but the result holds for an arbitrary collection of masses in an arbitrary external potential 
under rigid transport by noticing that the total potential must be of the form $V(Q-Q_0; \{r_j\})$, where $\{r_j\}$ represents a set of relative coordinates.    
The decomposition of $H_{\rm{com}}$  into terms for each ion, $H_{\rm{com}}= -\sum_i m_iq_i\ddot{Q}_0$, implies that ions of different mass should be subjected to different forces. However the available technology in linear Paul traps 
provides forces proportional to the charge (equal for all equally-charged ions), so the compensation is  a formal result without a feasible 
experimental counterpart.  
\begin{figure}[t!]
\begin{center}
\includegraphics[height=2.6cm]{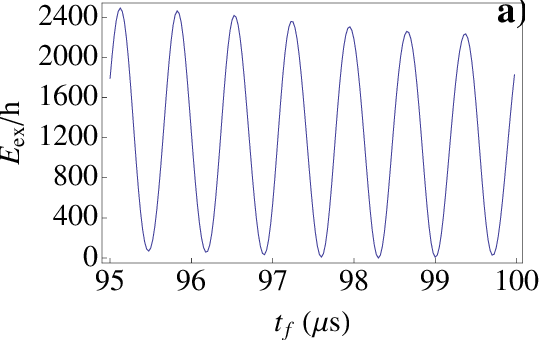}
\includegraphics[height=2.7cm]{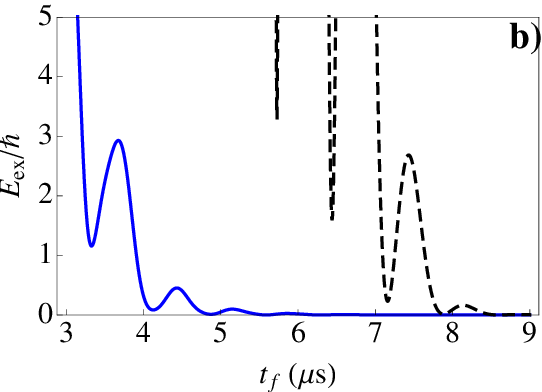}
\caption{\label{99}(Color online) 
{Excitation energy vs. final time for a) a linear-in-time transport of two ions,  $Q_0(t)=td/t_f$ and b) the trap trajectory designed in Eq.  (\ref{goodtrajectory}), (blue-solid line) and an ``error function'' trap trajectory, Eq. (\ref{gaussiantraj}) (black-dashed). We find optimal results for $\sigma=10^{-6} s$. 
Other parameters as in Fig. 1.}  }
\end{center}
\end{figure}

As for the linear displacement of the trap, $Q_0(t)=td/t_f$ in $[0,t]$, and at rest otherwise, 
we have performed numerical calculations of the final excitation energy
for different values of $t_f$ and the two ions considered in Sec. \ref{2ionsdm}. The excitation oscillates rapidly, 
see Fig. \ref{99} (a),
and the upper envelope reaches $0.1$ vibrational quanta of ion 1 for times as large as  $9.5$ ms. The first 
excitation minimum  with significant excitation reduction is around 
99 $\mu$s, see Fig. \ref{99} (a). Excitation minima  occur for each mode $\nu$ as zeroes of the Fourier transform of $\dot{Q_0}$ at $\Omega_\nu$ \cite{Reichle,David,Bowler}. For a linear-in-time trap displacement this occurs  every mode period. 
$99$ $\mu$s is a time when the transform of both modes vanishes. This excitation minimum, however, is very unstable with respect to 
small timing errors. In any case it is about twenty times larger than the times achieved in Sec. \ref{2ionsdm}). 

{
Finally, we compare the performance of our protocol in Eq. (\ref{goodtrajectory}) with an error-function trajectory
\cite{Reichle}. Imposing a Gaussian form on the velocity $\dot{Q}_0$ gives  
\beq\label{gaussiantraj}
Q_0(t)=-\frac{d}{2}\frac{\erf \left(\frac{-2t+t_f}{2\sqrt{2}\sigma}\right)}{\erf \left(\frac{t_f}{2\sqrt{2}\sigma}\right)}+\frac{d}{2},
\eeq
where $\sigma$ is the width of the Gaussian. In Fig. \ref{99} (b) we optimize $\sigma$ and compare the excitation for this  trajectory 
with the one in Eq. (\ref{goodtrajectory}). The error-function trajectory is clearly a good design, but still, the protocol developed in this paper outperforms it by a factor of two.}

%
%

In summary, we have described protocols for diabatic transport of mixed-species chains of ions that displace the minimum of a harmonic external potential along prescribed trajectories. Our protocols should allow for diabatic transport over distances and durations that are relevant for quantum information processing with minimal final excitation of the ion crystals. In past experiments on scalable  quantum information processing, adiabatic transport of mixed-species ion chains has been one of the most time consuming processes \cite{Home}, therefore the approaches described might lead to considerable practical improvements.  
Our work may be extended in several directions, e.g., to include noise, parameter drifts \cite{BEC,noise}  and anharmonicities \cite{Schulz,Li, Palmero}, or to optimize the trap trajectories according to different criteria \cite{Li}.

\acknowledgements{We acknowledge funding by Basque Country Government (Grant No. IT472-10, Ministerio de Econom\'{i}a y Competitividad (Grant No.
FIS2012-36673-C03-01), and the program UFI 11/55. M. P. acknowledges a fellowship by UPV/EHU. R. B., J. P. G. and D. L. are supported by IARPA under ARO Contract No. DNI-017389, ONR, and the NIST Quantum Information program.}

\end{document}